\documentclass{Source/egpubl}
\usepackage{Source/eg2024}
\usepackage{amssymb}
\usepackage{multirow}
\usepackage[normalem]{ulem}

\CGFStandardLicense

\usepackage[T1]{fontenc}
\usepackage{Source/dfadobe}  

\usepackage{macros}

\usepackage{cite}  
\biberVersion
\BibtexOrBiblatex
\electronicVersion
\PrintedOrElectronic
\ifpdf \usepackage[pdftex]{graphicx} \pdfcompresslevel=9
\else \usepackage[dvips]{graphicx} \fi

\usepackage{Source/egweblnk} 
\usepackage{lineno}

\usepackage{enumitem} 

\title[Predicting Perceived Gloss: Do Weak Labels Suffice?]%
      {Predicting Perceived Gloss: Do Weak Labels Suffice?}
\author[Guerrero-Viu et al.]
{\parbox{\textwidth}{
\centering 
Julia Guerrero-Viu$^{1*\mathsection}$
\orcid{0000-0002-2077-683X}, J. Daniel Subias$^{1*}$\orcid{0000-0002-5480-7462}, Ana Serrano$^{1}$\orcid{0000-0002-7796-3177}, Katherine R. Storrs$^{2}$\orcid{0000-0001-9573-8654}, Roland W. Fleming$^{3,4}$\orcid{0000-0001-5033-5069}, 
\\ Belen Masia$^{1}$\orcid{0000-0003-0060-7278} \& Diego Gutierrez$^{1}$\orcid{0000-0002-7503-7022}
        }
        \\
{\parbox{\textwidth}{\centering $^1$Universidad de Zaragoza, I3A, Spain
        \hspace{2mm}
         $^2$University of Auckland, New Zealand
         \hspace{2mm}
        $^3$Justus Liebig University Giessen, Germany \\
        $^4$Center for Mind, Brain and Behavior, Universities of Marburg and Giessen, Germany \\
        $^*$Joint first authors
        \hspace{2mm}
        $\mathsection$Corresponding author: juliagviu@unizar.es}
}
}

\begin{document}

\teaser{
 \includegraphics[width=\linewidth]{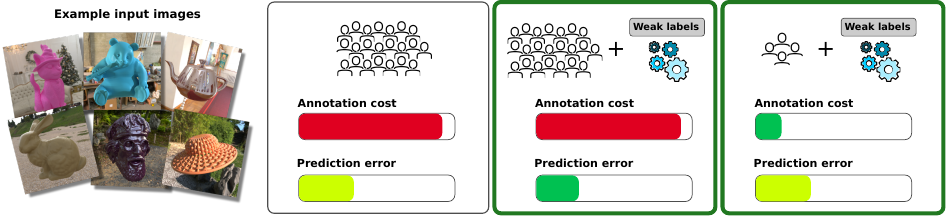}
 \centering
  \caption{We explore the benefits of weakly supervised learning to predict perceived gloss from images. 
  Commonly used supervised models require costly manual annotations of large datasets (left).
  \nw{In contrast, we propose to leverage simple weak labels, which can be automatically computed, achieving a twofold contribution: (1) our weak labels can be effectively combined with such costly annotations to reduce gloss prediction error, if accuracy is the priority (center); and (2) they enable a significant reduction in manual annotations of up to 80\% without sacrificing accuracy, if efficiency is the priority (right).}
  }
\label{fig:teaser}
}

\maketitle
\begin{abstract}
Estimating perceptual attributes of materials directly from images is a challenging task due to their complex, not fully-understood interactions with external factors, such as geometry and lighting. Supervised deep learning models have recently been shown to outperform traditional approaches, but rely on large datasets of human-annotated images for accurate perception predictions. Obtaining reliable annotations is a costly endeavor, aggravated by the limited ability of these models to generalise to different aspects of appearance. In this work, we show how a much smaller set of human annotations (``strong labels'') can be effectively augmented with automatically derived ``weak labels'' in the context of learning a low-dimensional image-computable gloss metric. We evaluate three alternative weak labels for predicting human gloss perception from limited annotated data. Incorporating weak labels enhances our gloss prediction beyond the current state of the art. Moreover, it enables a substantial reduction in human annotation costs without sacrificing accuracy, whether working with rendered images or real photographs.

\end{abstract}  
\section{Introduction}

The advent of powerful generative deep learning models for image synthesis and editing creates new opportunities, as well as new challenges, for computer graphics. In this context, techniques to navigate the massively high-dimensional latent spaces of such models in ways that are perceptually meaningful to humans become increasingly vital. Within the field of computer graphics, understanding and modeling human perception of material appearance has long been a fundamental challenge. This understanding is essential for accurately simulating the visual aspects of the physical world as perceived by humans and it has potential applications ranging from design and virtual prototyping to image editing and fabrication processes~\cite{chen2022gloss, subias2023wild}. In particular, gloss is one of the most perceptually salient appearance attributes~\cite{chadwick15VisRes, fleming2017material}. Unfortunately, how glossy a surface appears to a human viewer is a product of complex and only partially understood interactions between surface reflectance, illumination, and geometry~\cite{fleming2014visual,lagunas2021joint}. Perceived gloss, despite being grounded in photogeometric features such as the properties of reflected highlights~\cite{marlow12sciencedir}, cannot be well captured by either linguistic descriptors or objective measures, such as surface reflectance. This makes gloss an excellent candidate for a learned perceptual metric. 

\nw{Recently, it has been shown that complex representations of visual inputs are likely needed to capture perception of gloss in images~\cite{delanoy20SIG,fleming19COBS,lagunas2021joint}, drawing attention to deep learning approaches. 
Gloss perception is a high-dimensional problem that cannot be easily modeled by combining only physical material properties like roughness or specularity~\cite{fleming2014visual,chadwick15VisRes}. Such a model could not, for instance, account for the effects of shape or illumination on perceived gloss~\cite{marlow12sciencedir,storrs2021unsupervised,serrano2021effect}. Simple image statistics also fall short, due to the importance of spatial information (e.g., congruence of highlights, lowlights, and shading). Consequently, deep learning offers a framework for computing suitably complex visual features from which perceptual gloss metrics may be directly learning from detailed human annotations~\cite{lagunas2019similarity,serrano2021effect}.
}
State-of-the-art gloss predictions have been achieved using supervised methods, as demonstrated by Serrano et al.~\shortcite{serrano2021effect}. However, these approaches heavily rely on a substantial amount of human-annotated images, which can be prohibitively expensive to obtain. For instance, Serrano et al.~\shortcite{serrano2021effect} collected over 200,000 gloss ratings from more than 3,000 human participants for their dataset. Alternatively, methods employing fully unsupervised models, such as the one proposed by Storrs et al.~\shortcite{storrs2021unsupervised}, face challenges in capturing material properties in complex stimuli and realistic images. Object appearance in realistic images is influenced by multiple factors, including geometry, lighting, and viewpoint, creating a prohibitively high-dimensional space to learn in a fully unsupervised fashion from limited training data. Moreover, even after training, unsupervised models require further adjustments to align their latent dimensions with human perception.

In this work, we show that costly human annotations (\emph{strong} labels) can be effectively combined with automatically derived \emph{weak} labels in the context of learning a low-dimensional image-computable gloss metric.
First, we demonstrate that our weakly supervised approach achieves higher gloss prediction accuracy than previous supervised methods (Figure~\ref{fig:teaser}, center). It is not trivial that such an approach should work -- noisy weak labels can in some domains ``dilute'' high-quality strong labels, reducing a model's performance even as more training data is provided~\cite{li2019towards}.
Then, we leverage our weak labels to \emph{reduce} the annotation cost, allowing for an 80\% reduction in the amount of human-labeled data needed, without losing accuracy (Figure~\ref{fig:teaser}, right). In addition, we identify the need for a \emph{controlled} dataset to systematically evaluate gloss prediction models, as opposed to crowd-sourced existing datasets. For this purpose, we create a new test dataset that includes systematic variations in rotation, geometry complexity, illumination and specularity, and covers a reasonable and balanced range of glossiness levels. This new test dataset is annotated under constant, controlled viewing conditions that enable a \emph{reliable} evaluation of gloss prediction models. Finally, we show how our weakly supervised gloss predictor\nw{s are} consistent with human perception for such systematic variations and are able to generalize reasonably to ``in-the-wild'' real images.

In summary, we present the following contributions:
\begin{itemize}
    \item \nw{A study of three possible automatic weak labels (simple and cheap to obtain) that can effectively be leveraged to predict human perception of gloss without the need for a large amount of costly annotated data.}
    \item A test dataset with controlled variations and reliable annotations to evaluate perceptual gloss prediction models.
    \item \nw{An accurate gloss predictor} that outperforms previous methods, is consistent across changes in object view or illumination, and generalizes to real-world photographs.
\end{itemize}

Our datasets and gloss prediction models, as well as the training and evaluation code, are available in \url{https://graphics.unizar.es/projects/perceived_gloss_2024/}.

\section{Related Work}
\nw{Weakly supervised learning~\cite{zhou18_surveyWSL} has been widely applied in computer vision and graphics in several domains such as image classification~\cite{xiao2015_noisyWSL,cheng2020weakly}, object detection~\cite{ren2020instance,chen2020slv}, semantic segmentation~\cite{bearman2016_ssWSL,luo2020_dualbranch,kim2023_devilWSL}, or image retargeting~\cite{cho2017weakly}. However, in the fields of material appearance and perception science, contemporary studies still rely on tedious and expensive human annotations~\cite{lagunas2019similarity,delanoy2022generative}, which are the standard in psychophysical experiments. In this paper, we perform the first study of weakly supervised learning focusing on the field of material appearance, particularly on perceptual gloss prediction.}

\subsection{Gloss Metrics}

Finding an objective measure of gloss is a long-standing problem in computer graphics, industry, perception research and related fields. Since the work from Hunter et al.~\shortcite{hunter37NBS}, it has been recognized that a single physical measure is not sufficient to quantify physical gloss. An active field of research on the perceptual aspects of gloss has demonstrated that perceived gloss depends not only on surface reflectance properties but also on geometry, illumination, and motion~\cite{fleming01,fleming03JoV,pellacini00SIG,wills09ACMTG}. Chadwick et al.~\shortcite{chadwick15VisRes} provides a review of the study of gloss perception and measurement. 

One intuitive approach to quantifying gloss would be to consider a physical parameter (e.g., roughness) from an analytical BSDF model, as an objective measurement. However, such parameters of analytical models do not correlate well with perceived gloss~\cite{ngan06SR, wills09ACMTG}. Moreover, it is difficult to adjust BSDF parameters to approximate real-world measured materials while maintaining their perceptual quality~\cite{lavoue21CGF}. Alternatively, there are works showing that certain image statistics~\cite{motoyoshi07nature}, or visual cues~\cite{marlow12sciencedir} correlate with perceptual judgments of gloss, at least for simple stimuli. Last, the work of Westlund and Meyer~\shortcite{westlund01SIG} explores the objective measurement of physical gloss on analytical BSDF models following industry standards~\cite{hunter39ASTM}.

Following each of these proposals, we explore three different objective metrics to label gloss perception of rendered images, based on BSDF parameters, image statistics, and industry standards, and analyze their usefulness as weak labels to guide learning-based algorithms for gloss prediction.

\subsection{Dimensionality Reduction for Material Appearance}
Reducing the dimensionality of material representations while capturing the statistical structure of their appearance (either across images or BSDF tabulated data) is a challenging task with multiple applications for computer graphics, such as database searching and visualization~\cite{lagunas2019similarity}, gamut mapping~\cite{sun2017attribute} or material editing~\cite{serrano2021effect,delanoy2022generative, subias2023wild}.

Some early works show that linear reduction methods such as Principal Component Analysis (PCA) are good for finding directions that capture some traits of material appearance~\cite{matusik2003data,nielsen15ACMTG,serrano2016intuitive}. 
\rev{ Since the emergence of deep learning, supervised learning methods have demonstrated good performance for efficient compression of complex material representations, such as Bidirectional Texture Functions (BTFs)~\cite{rainer2019neural,rainer2020unified} or tabulated BSDFs~\cite{hu2020deepbrdf}. Other works also show that supervised methods are efficient in compressing images into low-dimensional representations while capturing the statistical structure of materials~\cite{lagunas2019similarity,serrano2021effect}.  However, these data-hungry algorithms require collecting large amounts of labeled data for training.
} On the other hand, recent works suggest that unsupervised learning methods are also capable of compacting high-dimensional BSDF tabulated data in low-dimensional latent vectors~\cite{benamira2022interpretable,zheng2021compact}, and keeping certain structures related to high-level perceptual attributes on simple image stimuli, such as gloss~\cite{storrs2021unsupervised} or translucency~\cite{liao2023unsupervised}. However, aligning such latent dimensions with human perception for the case of complex stimuli remains challenging.

In our work, we propose a weakly supervised encoder that compresses images in a low-dimensional latent space capturing the structure of gloss. Our weakly supervised learning approach allows us to reduce the annotation cost, by automatically labeling training images, while still supervising the learning process to simplify the task for complex stimuli.

\subsection{Material Appearance Datasets}

Existing databases of materials, such as MERL~\cite{matusik2003data}, RGL~\cite{Dupuy2018RGL} and UTIA~\cite{filip2014UTIA} contain measured BSDFs of different material types (e.g., metals, plastics or fabrics) and are widely used for material appearance applications~\cite{serrano2016intuitive, sun2017attribute, soler2018versatile}. To capture the complex interactions of materials with other factors that influence human perception of appearance, like geometry or illumination~\cite{lagunas2021joint}, previous works rely on image-based material datasets, such as CURet~\cite{dana1999reflectance}, OpenSurfaces~\cite{bell2013opensurfaces} or Lagunas et al.~\shortcite{lagunas2019similarity}. These large datasets are usually  annotated manually through crowd-sourcing experiments, in order to obtain ground-truth labels of perceptual attributes for material appearance, such as glossiness or metallicness. 

Our training process relies on the dataset from Serrano et al.~\shortcite{serrano2021effect}, which includes thousands of rendered images using a variety of real-world geometries, illuminations and measured materials, manually annotated with six perceptual attributes including glossiness. We further extend this dataset with new images of analytical materials, automatically labeled with our weak labels. Although massive crowd-sourced datasets are very useful to train learning-based models, they are not ideal for a systematic evaluation of gloss prediction models, since they can be noisy and do not cover controlled variations in factors affecting gloss perception. We thus create a new, controlled test dataset for evaluation, made up of 310 reliably annotated images.

\section{Datasets}
\label{sec:dataset}

We first describe the image dataset that we use to train our model (Section~\ref{subsec:training_dataset}). Then, we describe our new controlled test dataset to systematically evaluate gloss prediction results (Section~\ref{subsec:test_dataset}). \supp{We include our full test dataset in the supplemental material.} 

\begin{figure*}
    \centering
    \includegraphics[width=\textwidth]{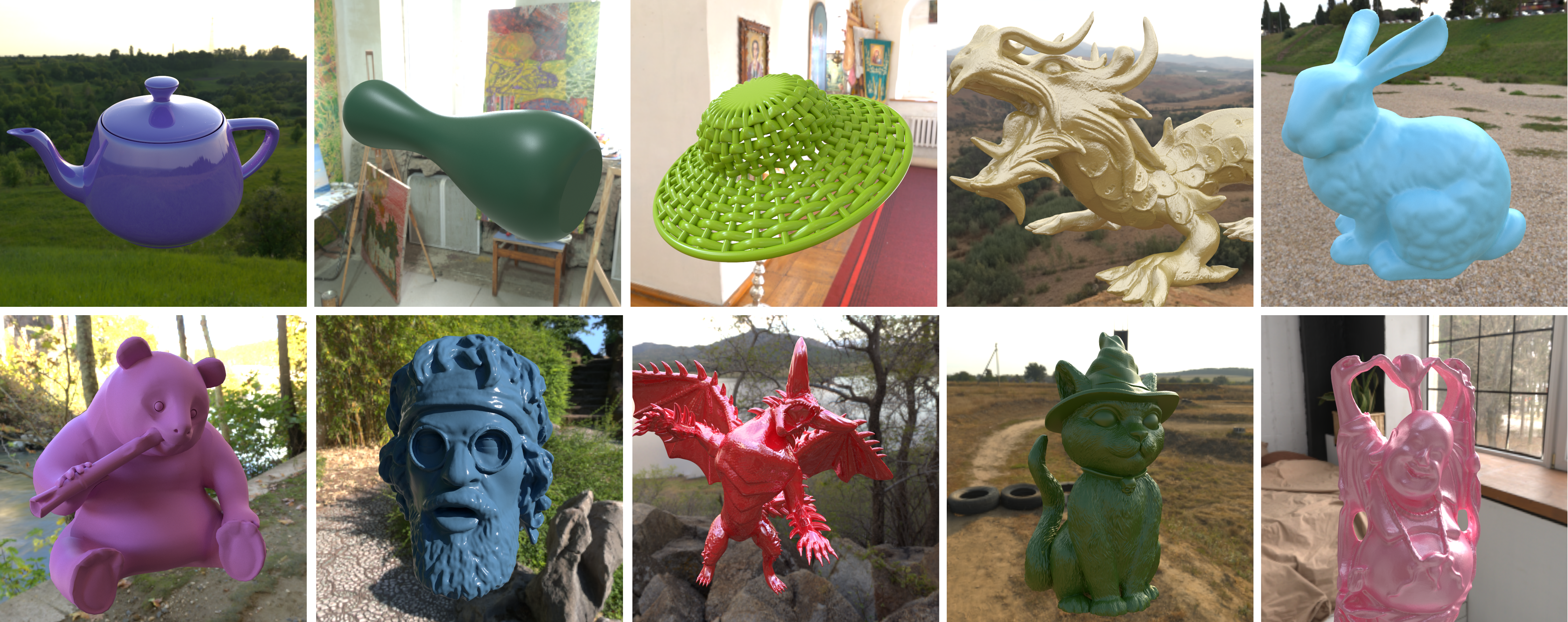}
    \caption{Example images from our training dataset. We extend the Serrano dataset~\cite{serrano2021effect} with a set of new images of varied geometries, illuminations and analytical materials using Disney's \textit{Principled BSDF}.
    }
    \label{fig:training_dataset}
\end{figure*}

\subsection{Training Dataset}
\label{subsec:training_dataset}

We use a large image-based material dataset to train our models. 
First, we rely on the dataset from Serrano et al.~\shortcite{serrano2021effect}, which includes a variety of measured materials under real-world geometries and illuminations. We use a subset of this dataset, after removing the geometries and materials used in our test dataset and all anisotropic materials, to simplify the range of material appearances, making a total of 23,616 images. Each of these images has an associated strong label of perceived glossiness, assigned manually by humans \nw{as a single rating in the 7-point Likert scale.}
\nw{Note that the commonly used Likert-scale ratings already capture potential non-linearities in perception}. In the following, for clarity, we always refer to this subset as the \emph{Serrano} dataset.

We then extend our training dataset with 38,250 \textit{new} images \nw{(never annotated by humans)} including new geometries, points of view, illuminations and analytical materials using Disney's \textit{Principled BSDF}~\cite{burley12SIG,burley15SIG}, for which we vary the values of the roughness $r$ and specular $s$ parameters in the intervals $[0, 0.5]$ and $[0.1, 5]$, respectively. The images were rendered using Mitsuba 3~\cite{Mitsuba3}, using a global gamma-exposure tone mapping operator to avoid introducing non-uniform contrast changes following previous work~\cite{serrano2021effect}. 
\nw{We annotate this extended dataset with our automatically-computed weak labels (see Section~\ref{sec:weak_labels}), with the purpose of offering an effective alternative to human labeling of a full dataset, improving generalization at a very low cost.}
Figure~\ref{fig:training_dataset} shows a representative subset\supp{, and additional details are included in the supplemental material}. 

\begin{figure}[t!]
    \centering
    \includegraphics[width=\linewidth]{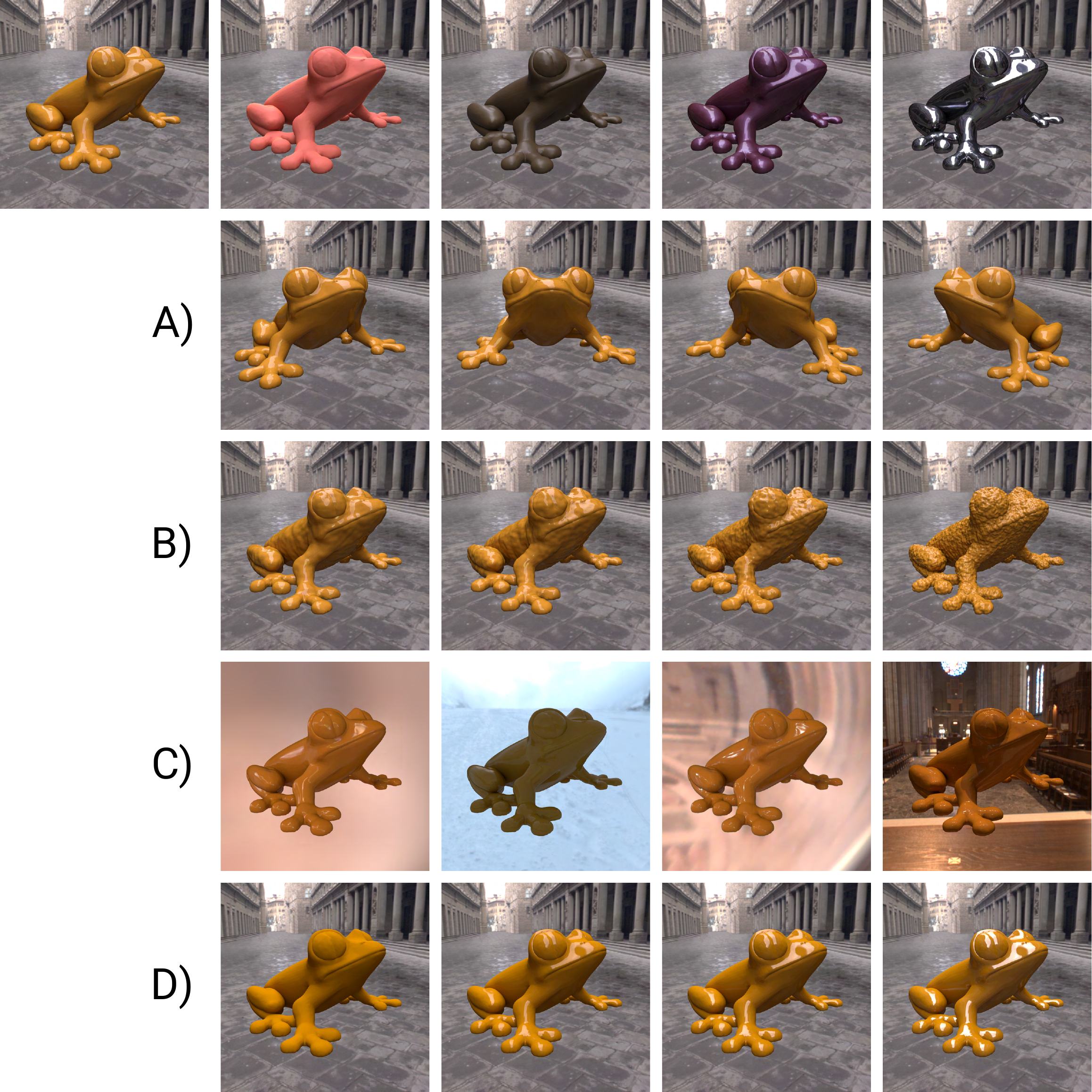}
    \caption{
    Representative images of our controlled test dataset. Top row shows our five baseline measured materials under the baseline geometry \emph{frog} and baseline illumination \emph{uffizi}. Rows A-D show individual variations for one example material: A) different rotations, B) different geometry complexity by increasing bumpiness of the surface, C) different illuminations, and D) different levels of the specular parameter for the analytical fitting of the material. 
    }
    \label{fig:test_dataset}
\end{figure}

\subsection{Test Dataset}
\label{subsec:test_dataset}

Several datasets exist that pair images of varying material appearance with ground-truth data in the form of human judgements of gloss~\cite{serrano2021effect, serrano2016intuitive, delanoy2022generative, storrs2021unsupervised}. Most of these datasets are built with the goal of training, testing or calibrating models that mimic human perception of gloss, and feature varying material properties, lighting conditions, geometries and camera viewpoints. However, they are not ideal for a systematic evaluation of gloss prediction models, for several reasons. First and foremost, images in the dataset do not include \textit{controlled} variations, such as systematic changes in point of view, or illumination frequency, to measure consistency. 
Second, annotations are given through large crowd-sourced experiments under varying, uncontrolled viewing conditions, and thus can be noisy and unreliable for specific images. And last, annotations might be highly unbalanced with respect to levels of perceived gloss, which can lead to erroneous conclusions when reporting an average error metric (e.g., a constant baseline model that always predicts the lowest gloss level for every image could get a low average error if the majority of the images in the test dataset are completely matte).
Therefore, we create a \emph{test dataset}, annotated under constant, controlled viewing conditions, which we use to reliably evaluate our models. 

Our test dataset includes twenty baseline samples, from which we generate controlled variations, yielding a total of 310 images. The baseline samples are created through combinations of geometry, material (measured BSDFs), and illumination (environment maps), and cover a reasonable range of appearances (see Figure~\ref{fig:test_dataset}, top row). Specifically, our baseline includes:

\begin{itemize} 
    \item Two geometries with different complexity: \emph{frog} and \emph{bumpy\_sphere}.
    \item Two illuminations with different distributions of spatial frequencies: \emph{uffizi} and \emph{st\_peters}.
    \item Five measured materials, from the MERL database~\cite{matusik2003data}, with different glossiness: \emph{pink\_plastic}, \emph{fruitwood}, \emph{violet\_acrylic}, \emph{specular\_yellow\_phenolic} and \emph{aluminium}.
\end{itemize}

For each combination of the baseline materials, illuminations and geometries, the following variations are generated (see Figure~\ref{fig:test_dataset}, rows A through D):

\begin{enumerate}[label=\Alph*)]
    \item Five different rotations of the object.
    \item Five different levels of geometry complexity, by increasing the bumpiness of the surface.
    \item Three additional illuminations. Ranked in order of increasing high-frequency content~\cite{brossier2004real}, the full set of illuminations is: \emph{st\_peters\_blurred}, \emph{glacier}, \emph{uffizi}, \emph{st\_peters}, and \emph{grace}. We obtain \emph{st\_peters\_blurred} by applying a Gaussian filter of size $20x20$ to the original \emph{st\_peters}, to offer a direct comparison.
    \item Five different levels of the specular parameter for the Ward-Duer~\cite{dur2006ward} analytical fitting of the material, sampling the range [-0.05, +0.05]. \rev{Although the Ward-Duer model has limitations modeling the Fresnel effects, this model adequately fits our need to vary the specular level in the MERL database materials, as the parameters for the material fitting are publicly available in the work of Ngan et al.~\cite{ngan06SR}.}  
\end{enumerate}

Our resulting dataset includes challenging examples (e.g., non-conventional illumination maps like \emph{glacier}), both measured and analytical materials, and large differences in geometry complexity. As well as providing a challenging benchmark for gloss prediction, these variations provide a means to evaluate gloss constancy, both for human annotators and for models. 

Each of these 310 images is manually annotated for gloss, on a 7-point Likert scale, by five different subjects 
\nw{(ages 24-35, three females and two males, all claiming experience in computer graphics).} 
Annotation is done under constant, controlled viewing conditions (\nw{SDR} display and \nw{fixed} lighting).  
Agreement between annotators is high (Krippendorff’s alpha~\shortcite{krippendorff2011computing} $= 0.87$; $1.0$ would indicate perfect agreement), and annotations are well balanced between gloss levels (see further details in the supplemental material). 
\section{Gloss Prediction with Weak Labels}
\label{sec:training_details}

To obviate the need to collect a large number of strong (manual) gloss labels via time-consuming user studies, we propose combining just a small set of strong labels with a larger set of \textit{automatically computed} weak labels. 

\subsection{Automatic Weak Labels}
\label{sec:weak_labels}
Our objective in developing automatic weak labels is to ensure their ease of computation while maintaining a rough correlation with the perception of gloss. These labels are obtained through simple and coarse approximations. To illustrate this, we explore three simple methods to automatically produce weak labels for our images, based on i) Disney's \textit{Principled BSDF} model~\cite{burley12SIG,burley15SIG}; ii) image statistics; and iii) industry metrics~\cite{hunter39ASTM}, respectively. As a result, every image $\mathbf{x}$ in our analytical dataset has three different weak labels $\{y_0, y_1, y_2\}$ \nw{(in the range $[1,7]$)} associated with it. 

\paragraph*{Disney's Principled BSDF Model ($y_0$)}
\nw{Previous works~\cite{pellacini00SIG,wills09ACMTG,adams2018naturally} suggest that roughness $r$ and specular $s$ are the two main physical parameters related to gloss perception.
}
We use a weighted combination of these parameters \nw{in Disney's Principled BSDF Model} to approximate the gloss level for each image, as $y_0 = \left\lfloor \lambda_s s + \beta (\alpha -\lambda_r r^2 ) \right\rfloor$ where $\lambda_s$ and $\lambda_r$ are weights to balance the distribution of the labels, which we empirically set to 0.95 and 1.2 respectively, and $\beta$ and $\alpha$ are scalars to translate the roughness term to the interval $[1,2]$ (set to 4 and 0.5, respectively). 

\paragraph*{Image Statistics ($y_1$)} Some works point to simple image statistics as indicators of human perception of gloss for simple stimuli~\cite{motoyoshi07nature,sharan08JOSA,wiebel15VR}.
Although such approaches may not generalize well to complex images or high-frequency illumination, we analyze their usefulness as weak labels since they are easy to compute and not linked to any specific BSDF model.
We define $y_1 = \frac{1}{N} \sum_{i=1}^{N} (x_i- \overline{x})^3 / \sigma^3$ as the skewness of the luminance histogram~\cite{motoyoshi07nature}, where $\overline{x}$ and $\sigma$ are its mean and the standard deviation, respectively. We compute the skewness per geometry and material under low-frequency illumination in grey scale, where we empirically observe a reasonable correlation to gloss perception. \rev{To compute the skewness only from pixels inside the objects, we remove the background using a mask}. Then, we assign the same label $y_1$ across different illuminations. 
\nw{Recent literature shows that more complex visual features are better correlated to human perception~\cite{schmid2023material} but these require access to matte and specular components separately, so we choose skewness for its simplicity.}

\paragraph*{Industry Metrics ($y_2$)} The standard D523 from the American Society for Testing and Materials (ASTM) provides a procedure to measure gloss on real-world materials~\cite{hunter39ASTM}. Westlund and Meyer~\shortcite{westlund01SIG} apply its methodology to compute the 
ratio between the radiance given by simple BSDF models and an analytical approximation of a polished black glass. We follow a similar approach and calculate the log-ratio between the  radiance computed with Disney's \textit{Principled BSDF} (used to render our analytical dataset) and a black glass modeled by a GGX microfacet model ~\cite{walter07egsr}, as $y_2 =  \log(R_{d} + 1) / \log(R_{g}  + 1)$, where $R_{d}$ and $R_{g}$ are the radiance of  Disney's \textit{Principled BSDF} and the black glass respectively, computed at an angle of 20º with respect to the normal. Both $R_{d}$ and $R_{g}$ are computed with the radiance meter provided by Mitsuba 3~\cite{Mitsuba3}. 
\begin{figure}[t!]
    \centering
    \includegraphics[width=\linewidth]{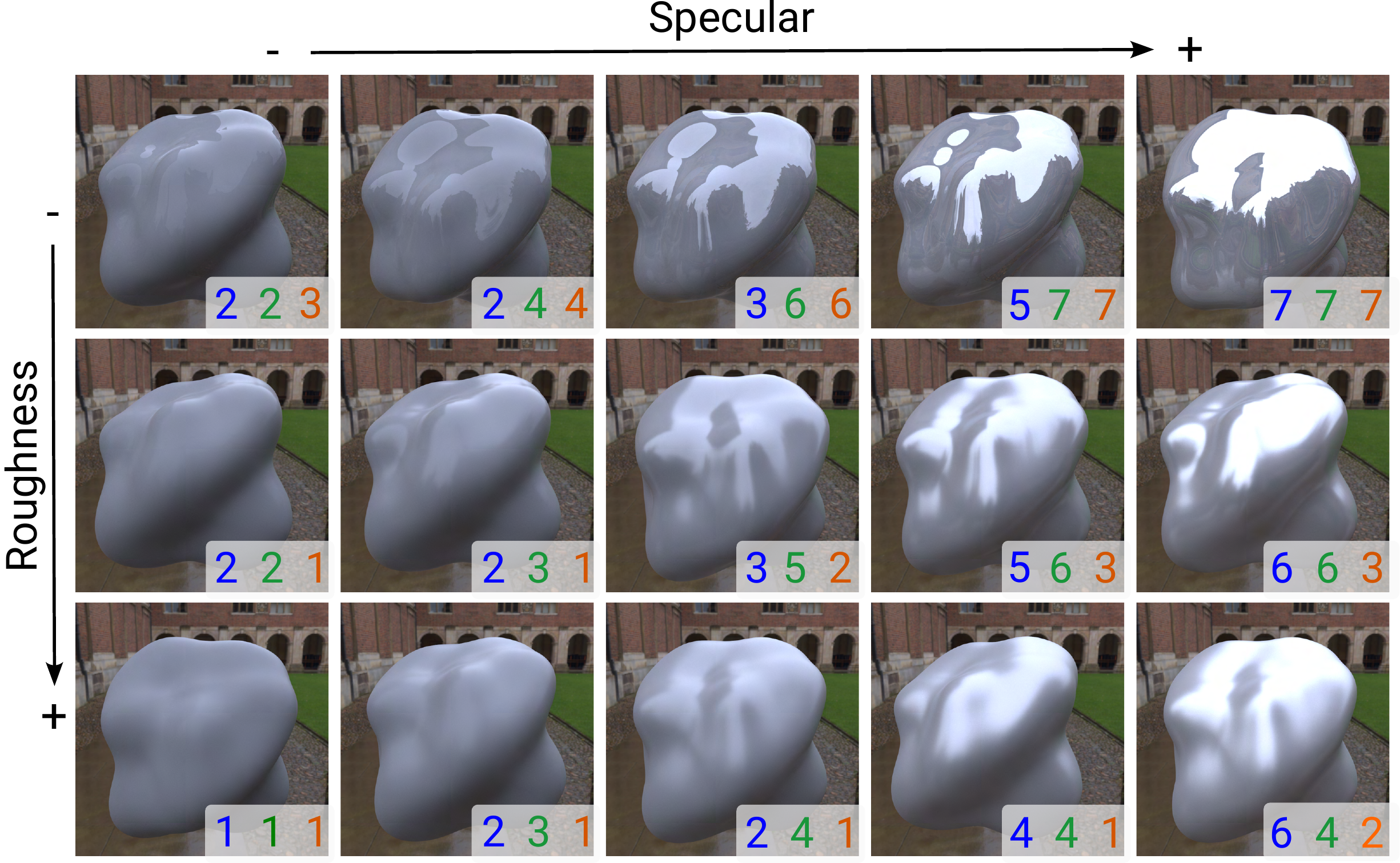}
    \caption{Weak labels automatically computed for example images in our training dataset: based on the BSDF model (blue), image statistics (green) and industry metrics (orange). We show samples from the \emph{blob} geometry under the \emph{cambridge} illumination and grey albedo with different combinations of roughness and specular parameters from Disney's \textit{Principled BSDF}. Although none of these labels are precise indicators of perceived gloss, they sufficiently correlate with gloss perception to be used as weak labels. \nw{All weak labels are in the range [1, 7].}}
    \label{fig:weak_labels}
\end{figure}
Each resulting $y_i$ approximates the gloss level of $\mathbf{x}$ on a 7-point Likert scale. We show examples of our resulting weak labels in Figure~\ref{fig:weak_labels}. Although none of them is suitable to directly derive a gloss metric (nor have they been designed for that purpose), we can observe how these labels reasonably correlate with perceptual gloss, which suffices for our purposes. 

\subsection{Gloss Predictor} We next describe the network architecture and our weakly supervised framework for training a gloss predictor, together with implementation details for reproducibility. 
During training, we use data augmentation to increase the number of images by flipping, cropping, shifting, rotating, scaling, and adding Gaussian and Poisson noise. 

As a feature extractor, following the work from Serrano et al.~\shortcite{serrano2021effect}, we use a VGG16 architecture and remove its last layer. Then, we initialize our feature extractor with weights from pre-training on the ImageNet dataset~\cite{deng09CVPR} and introduce two fully connected layers to compress the features into a reduced 20-dimensional latent vector $\textbf{z}$. In addition, since we are interested in generating a linearly separable space, we introduce a linear regression layer that predicts the gloss level from $\textbf{z}$, constrained to the range $[0,1]$.

We train our gloss predictor for regression by minimizing the Mean Absolute Error (MAE) between its predicted value $\hat{y}$ and the training label $y$ normalized \nw{(min-max normalization)} to the interval $[0,1]$. This training label can be either \emph{strong} (coming from human annotations and computed as the median between users) or \emph{weak} (automatically computed with any of our three methods explained in Section~\ref{sec:weak_labels}). Therefore, our training loss $L_{MAE}$ is defined as:
\begin{equation}
 L_{MAE} = \frac{1}{N} \sum_{j=1}^{N} |y_{j} - \hat{y_j}| ,
\label{eq:mae}
\end{equation} 
where $N$ is the number of images in the batch.
We train our predictor for $35$ epochs with batch size $N = 4$ and input resolution $512$x$512$, using the Ranger optimizer~\cite{zhang2019lookahead, liu2020variance} with an initial learning rate set to $10^{-5}$.

\section{Results}
\label{sec:results}
We evaluate our gloss predictor on our controlled test dataset described in Section~\ref{subsec:test_dataset}. We begin by thoroughly evaluating the impact of our weakly supervised learning strategy with our three types of weak labels (Section~\ref{subsec:weakly-supervised}),
\nw{followed by a consistency evaluation across the controlled variations in the test dataset (Section~\ref{subsec:consistency_evaluation}). Then, we compare the accuracy of our gloss predictors to the current state of the art (Section~\ref{subsec:sota}) and show the ability of our models to generalize to more diverse data, including real images (Section~\ref{subsec:generalization}). Finally, we perform ablation studies to validate the impact of our main design decisions (Section~\ref{subsec:ablation}), and explore the structure of our 20-dimensional latent space with respect to human perception of gloss (Section~\ref{subsec:latent_space}).}

\begin{table}
\centering
 \caption{Mean Absolute Error (MAE $\downarrow$
 ) on our controlled test dataset for our gloss predictors trained on the following data: using only strong human labels (with the 100\% of the Serrano dataset or with only $20\%$ of it, S. only), 
 combining these strong labels with our weak labels based on either BSDF model (S.+BSDF), image statistics (S.+Image stats.) or industry metrics (S.+Industry) \rev{and using only our weak labels (S. $0\%$)  
  }
 We use S. to refer to the Serrano dataset.
}
 \small
\begin{tabular}{r|c|c|c|c}

  & S. only & S.+BSDF & S.+Image stats. & S.+Industry \\ \hline 
 S. 100\% & $0.1510$ & $\textbf{0.1207}$ & $0.1389$ & $0.1484$  \\
 S. 20\% & $0.2091$  & $0.1538$ & $0.1550$ & $0.1797$ \\ 
\rev{S. 0\%} & \rev{-} & \rev{$0.3114$} & \rev{$0.3015$} & \rev{$0.3466$}  \\ \hline

\end{tabular}
 \label{tab:all_weakly_results}
\end{table}
\begin{figure*}[t!]
    \centering
    \includegraphics[width=\textwidth]{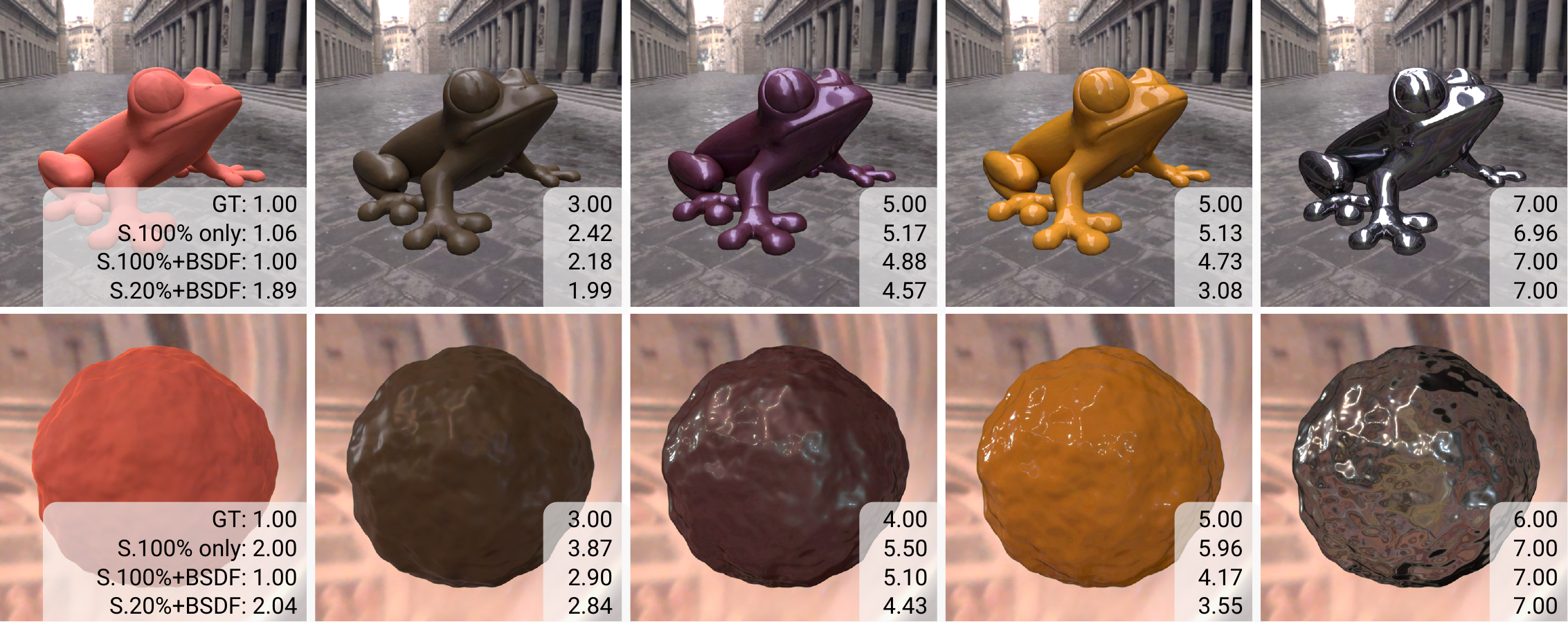}
    \caption{
    Qualitative results of our gloss predictor\nw{s} on example images from our controlled test dataset. 
    \nw{The numbers in the insets indicate, for every input image, the ground-truth judgement (GT) and the predictions from our models (from top to bottom): supervised model trained only with strong labels (S.100\% only, second line), weakly supervised model with strong labels combined with our BSDF weak labels (S.100\% + BSDF, third line) and weakly supervised model with a subset of the strong labels combined with our BSDF weak labels (S.20\% + BSDF, fourth line).}
    All gloss ratings are in the range $[1, 7]$.
    }
    \label{fig:qualitative}
\end{figure*}

\subsection{Evaluation Metrics}
To quantitatively evaluate our predictions, we follow recent related work~\cite{serrano2021effect} and use the Mean Absolute Error (MAE), which measures the absolute distance to the ground-truth gloss judgements of our controlled test dataset, normalized to the interval $[0, 1]$. We define the ground-truth (GT) judgement per image as the median across our five human ratings, as a robust estimator of human perception. By considering the median, we obtain a more reliable estimation that is less influenced by individual variations or extreme judgments. In addition, in order to evaluate the consistency of our gloss predictions and their rank performance with respect to the ground truth, we compute the Pearson correlation, which measures the linear dependence between them without taking into account the absolute scale. Finally, to consider monotonic, but not necessarily linear, relationships we also compute Spearman's rank correlation. Both correlation coefficients are bounded to the range $[-1, 1]$, where $1$ means perfect positive correlation.

\subsection{Weakly Supervised Learning}
\label{subsec:weakly-supervised}

\nw{\textbf{\textit{``Same annotation cost, better performance'':}}}
We show the effectiveness of our weak labels in improving gloss predictions without the need to collect a large amount of costly manual annotations. In Table~\ref{tab:all_weakly_results} (first row) we show MAE results on our controlled test dataset for our gloss predictor trained only on the \emph{Serrano} dataset with strong labels, compared to the same predictor trained on our full training dataset, including our weak labels. Although this is a challenging task, as noisy weak labels could easily degenerate learning performance even with more data~\cite{li2019towards}, our weakly supervised gloss predictor outperforms the predictor trained only with strong labels, across our three different weak labels.

\rev{In Table~\ref{tab:all_weakly_results} (third row) we show MAE results on our controlled test dataset featuring our gloss predictor trained exclusively using weak labels.}
Despite the limited ability of our weak labels \emph{alone} to capture human gloss perception accurately, they are still sufficiently good to guide the learning process when \emph{combined} with human-annotated data. Our weak labels allow us to train on a larger training dataset, including new geometries, illuminations, and materials, helping the model better generalize without incurring the high cost of manually annotating such new data. 

\nw{\textbf{\textit{``Same performance, substantially less annotation cost'':}}}
Next, we study to what extent our weak labels can help mitigate the cost of collecting large human-annotated datasets, while still maintaining performance. We train our gloss predictor on a small \nw{randomly selected} subset of the \emph{Serrano} dataset, consisting of only $20\%$ of the strongly labeled images. 
We combine this reduced human-annotated subset with our extended weakly labeled dataset and show that we get similar, competitive performance (Table~\ref{tab:all_weakly_results}, second row). In particular, for both BSDF and image statistics weak labels we can reduce the cost of human annotation by $80\%$, without losing accuracy with respect to using the 100\% of the \emph{Serrano} dataset \nw{(MAE = $0.15$ in S.20\%+BSDF, S.20\%+Image stats and S.100\% only)}. These two different weak labels are complementary: for synthetic images with analytical materials we could rely on labels based on the BSDF model, while image statistics could be easily computed from any simple image.
\nw{Hereafter, we will focus on our weakly supervised models using the BSDF weak labels (Table~\ref{tab:all_weakly_results}, S.100\%+BSDF and S.20\%+BSDF), as they are the ones that show the best performance. Refer to the supplemental material for further results using the image statistics and industry weak labels.} \\

We show qualitative results in Figure~\ref{fig:qualitative}.
\nw{We observe how incorporating our weak labels in addition to the 100\% of the strongly labeled \emph{Serrano} dataset (S.100\%+BSDF) leads to gloss predictions that better correlate with the ground-truth human judgements compared to training only on the strongly labeled data (S.100\% only). Additionally, combining our weak labels with a subset of the \emph{Serrano} dataset (S.20\%+BSDF) still maintains reasonably accurate gloss predictions.}
\supp{More qualitative results are included in the supplemental material.} 

\begin{table}
 \caption{Quantitative results when varying each of the confounding factors in our test dataset. We include results of our weakly supervised gloss predictor\nw{s, both trained with our BSDF weak labels jointly with the 100\% of the Serrano dataset (S.100\% + BSDF) vs. the 20\% of it (S.20\% + BSDF)}. We show MAE, Spearman and Pearson correlations results for every variation.}
\begin{tabular}{l|c|c|c}
Variation       & MAE $\downarrow$    & Spearman $\uparrow$       & Pearson $\uparrow$        \\ \hline \hline
\multicolumn{4}{c}{\nw{S.100\% + BSDF}}   \\ \hline
A) Rotation     & 0.1043  & 0.9050 & 0.9119 \\ 
B) Bumpiness        & 0.1076  & 0.8796 & 0.8979 \\
C) Illumination & 0.1004  & 0.9089 & 0.9129 \\ 
D) Specularity     & 0.1448  & 0.8501 & 0.9086 \\ 
 \hline
\multicolumn{4}{c}{\nw{S.20\% + BSDF}}  \\ \hline
\nw{A) Rotation}     & \nw{0.1381}  & \nw{0.8746} & \nw{0.8591} \\ 
\nw{B) Bumpiness}        & \nw{0.1460} & \nw{0.8262} & \nw{0.8248} \\
\nw{C) Illumination} & \nw{0.1356}  & \nw{0.8565} & \nw{0.8411} \\ 
\nw{D) Specularity}     & \nw{0.1801}  & \nw{0.8100} & \nw{0.8696} \\ 
 \hline
\end{tabular}
\label{tab:metrics_over_cofounding_factors}
\end{table}
\begin{figure*}[t!]
    \centering
    \includegraphics[width=\linewidth]{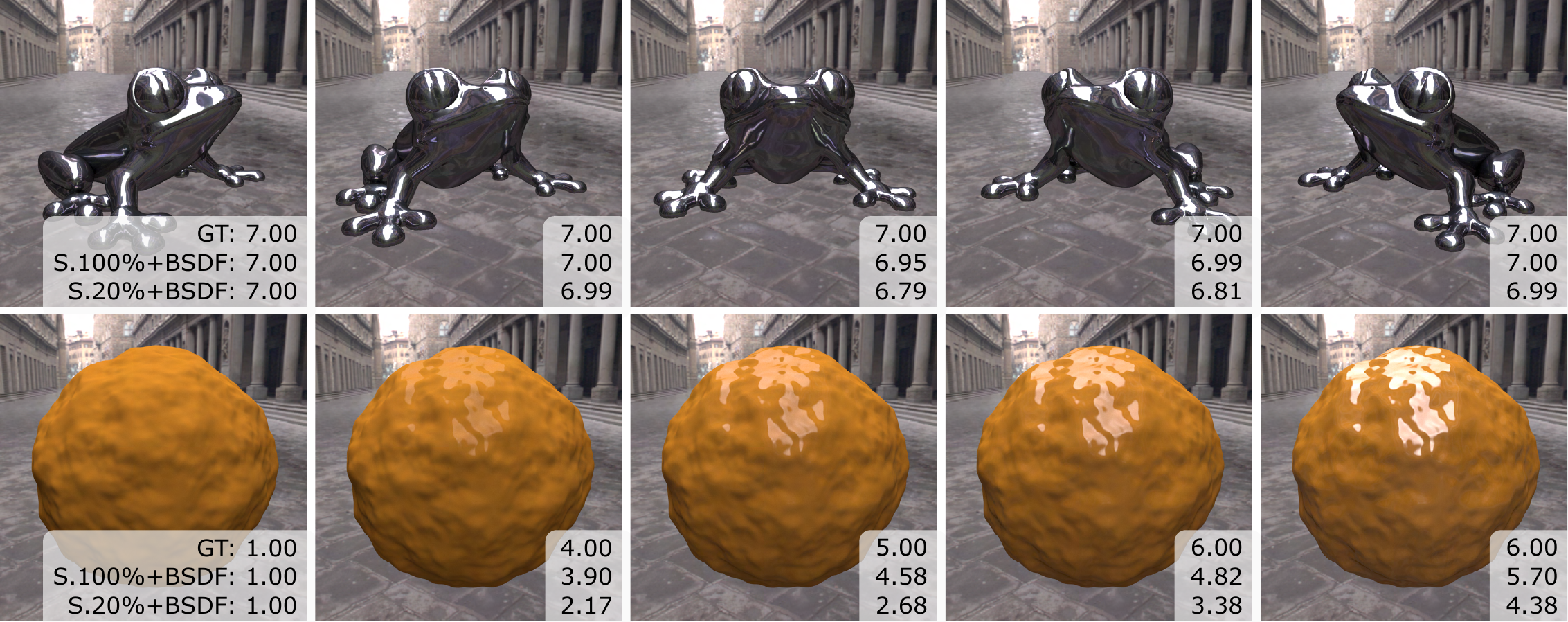}
    \caption{
    Qualitative results of our \nw{weakly supervised} gloss predictor\nw{s} when varying one confounding factor at a time on our test dataset. We show: variation across different rotations for the frog geometry with uffizi illumination and aluminium material (top), and variation across increasing specularity for the bumpy\_sphere geometry with uffizi illumination and the Ward-Duer BRDF fitting of specular\_yellow\_phenolic material (bottom). 
    \nw{The numbers in the insets indicate the ground-truth judgements (GT) and the gloss predictions from our weakly supervised models trained with BSDF weak labels jointly with the 100\% of the Serrano dataset (S.100\%+BSDF, second line) and the 20\% of the Serrano dataset (20\%+BSDF, third line)}. All gloss ratings are in the range $[1, 7]$.
    } 
    \label{fig:confounding_factors}
\end{figure*}

\subsection{Consistency Evaluation}
\label{subsec:consistency_evaluation}
We evaluate the consistency of our weakly supervised gloss predictors for every variation of different confounding factors present in our test dataset, both quantitatively and qualitatively. Table~\ref{tab:metrics_over_cofounding_factors} shows how our predictor\nw{s} perform consistently well across variations of rotation, bumpiness, illumination, and specularity, for the three different evaluation metrics. Although the error (MAE) is slightly higher when increasing the specular level (row D), the Pearson correlation remains high, indicating that our predictor\nw{s} successfully capture the trend of the perceived glossiness.

Figure~\ref{fig:confounding_factors} illustrates qualitative results for variations of rotation and specularity. The top row of images shows how human perception of gloss (GT) remains constant under different rotations of the object. Accurately, our weakly supervised predictions exhibit very low variation 
\nw{(std = $0.01$ and $0.09$, respectively)}, also preserving gloss constancy. The bottom row shows the effect of increasing the specular level; as expected, the ground-truth perceived gloss also increases (although not linearly). We observe how our predictor\nw{s} \nw{tend to} underestimate the absolute gloss predictions \nw{(slightly for the case of the S.100\%+BSDF model and more clearly for the S.20\%+BSDF one)}, yet effectively capture the increasing trend \nw{in both cases}. \supp{Further qualitative results for variations in bumpiness and illumination are included in the supplemental material.}

\subsection{Comparison to State of the Art}
\label{subsec:sota}

We compare our gloss predictors against a supervised state-of-the-art gloss predictor \nw{from Serrano et al.~\cite{serrano2021effect} (available in \url{https://github.com/Hans1984/material-illumination-geometry})}
on our controlled test dataset. Results are shown in Table~\ref{tab:comparison_sota}, where we see that \nw{all our predictors} match significantly better with ground-truth human judgements across all evaluation metrics, \nw{even when trained on a smaller amount of human-annotated data.}

\begin{table}
\centering
 \caption{Quantitative evaluation of our gloss predictors and Serrano et al.~\shortcite{serrano2021effect} gloss predictor 
 on our controlled test dataset. \nw{We include results of our weakly supervised gloss predictors, both trained with our BSDF weak labels jointly with the 100\% of the Serrano dataset (S.100\% + BSDF) vs. the 20\% of the Serrano dataset (S.20\% + BSDF)}. We show results for MAE, Spearman and Pearson correlations with respect to the ground truth.}
\begin{tabular}{r|r|c|c|c}
\multicolumn{2}{c|}{Gloss predictor} & MAE $\downarrow$  & Spearman $\uparrow$ & Pearson $\uparrow$ \\ \hline \hline
\multicolumn{2}{r|}{Serrano et al.} & 0.3293  & 0.5662 & 0.5358 \\ \hline
 \multirow{2}{*}{Ours} & S.100\%+BSDF  & \textbf{0.1207} & \textbf{0.8594} & \textbf{0.8788}  \\ 
 & \nw{S.20\%+BSDF}   & 0.1538 & 0.8366 & 0.8228  \\\hline
\end{tabular}
 \label{tab:comparison_sota}
\end{table}

\subsection{Generalization}
\label{subsec:generalization}

\begin{figure}[t!]
    \centering
    \includegraphics[width=\linewidth]{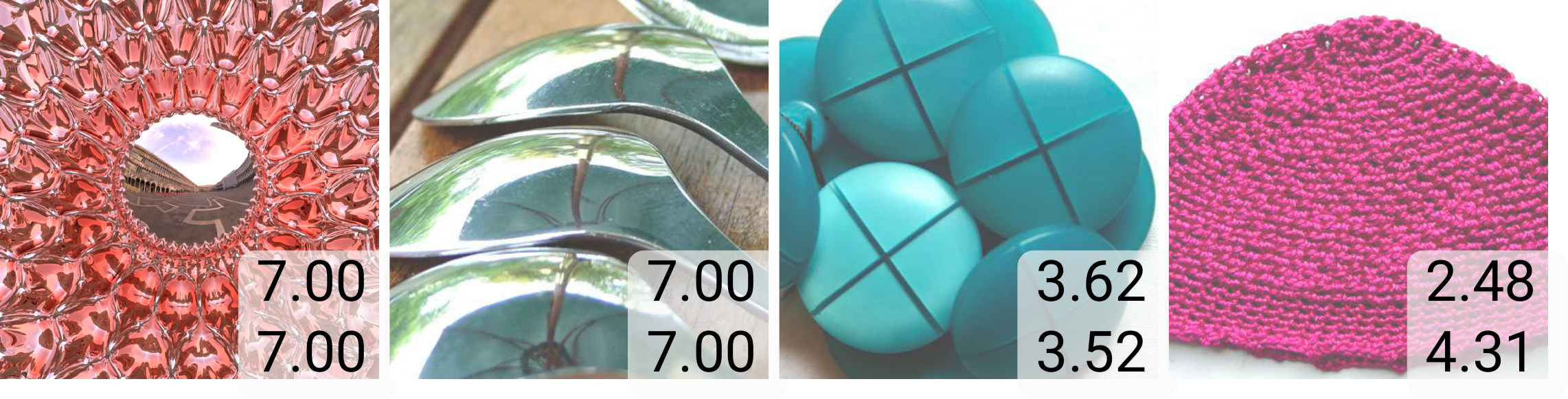}
    \caption{
    Qualitative results on challenging out-of-distribution images from test set B~\cite{serrano2021effect}, that include (from left to right): a synthetic image with a very glossy mirror-like surface and a highly complex geometry, several real objects with mirror-like surface, a real image of plastic material, and a real photograph of cloth with a knitted pattern. 
    \nw{The numbers in the insets indicate gloss predictions from our weakly supervised models trained with BSDF weak labels jointly with the 100\% of the Serrano dataset (S.100\%+BSDF, first row) and the 20\% of the Serrano dataset (S.20\%+BSDF, second row)}.
    Predictions are in the range $[1, 7]$.
    }
    \label{fig:generalization}
\end{figure}

Additionally, we evaluate how well our weakly supervised predictor\nw{s} generalize to challenging, out of distribution data, using the test set B from Serrano et al.~\shortcite{serrano2021effect}. This test set includes $78$ synthetic images depicting new materials, geometries and illuminations not included in our training data, as well as $20$ ``in-the-wild'' real photographs of materials from the Flickr Material Database (FMD)~\cite{sharan2014FMD}. We report quantitative results in Table~\ref{tab:generalization}, and show illustrative examples in Figure~\ref{fig:generalization}. As we can see,  
our gloss predictor\nw{s} obtain a reasonable performance despite the challenging out-of-distribution dataset: 
\nw{mirror-like surfaces (left and middle-left) are accurately predicted by our models as highly glossy despite their complex reflections; the cloth material (right) is predicted as medium-gloss by our S.20\%+BSDF model (second row), probably due to the high contrast of the knitted pattern, while our best model (S.100\%+BSDF, first row) is able to successfully predict it as a low-gloss object.}

\begin{table}
 \caption{Quantitative evaluation of the generalization capabilities of our weakly supervised gloss predictors and Serrano et al.~\shortcite{serrano2021effect} gloss predictor on their test set B, which contains real photographs. \nw{We include results of our weakly supervised gloss predictors, both trained with our BSDF weak labels jointly with the 100\% of the Serrano dataset (S.100\% + BSDF) vs. the 20\% of the Serrano dataset (S.20\% + BSDF)}. We show results for MAE, Spearman and Pearson correlations with respect to the ground truth.}
\begin{tabular}{r|r|c|c|c}
\multicolumn{2}{c|}{Gloss predictor} & MAE $\downarrow$ & Spearman $\uparrow$ & Pearson $\uparrow$ \\ \hline \hline
\multicolumn{2}{r|}{Serrano et al.} & 0.3327  & 0.4546 & 0.4266 \\ \hline
\multirow{2}{*}{Ours} & \nw{S.100\% + BSDF}   & \textbf{0.2236} & \textbf{0.6625} & \textbf{0.6570}  \\
 & \nw{S.20\% + BSDF}   & 0.2386 & 0.6208 & 0.6063  \\ 
 \hline
\end{tabular}
 \label{tab:generalization}
\end{table}

\subsection{Ablation Studies}
\label{subsec:ablation} 

We analyze how different design decisions of our training affect performance. We independently evaluate the possibility of removing our data augmentation pipeline \nw{and} adding background masks to the input images \nw{(masking out the environment by setting all background pixels to $0$)}.
As shown in Table~\ref{tab:ablation}, the data augmentation pipeline has a major positive effect on the predictions as it helps the model\nw{s} to  generalize better to unseen data. Background masks do not seem to help the model\nw{s} to estimate human gloss perception. We hypothesize that the network\nw{s} might leverage the contextual cues from the background to disambiguate between material properties and illumination, similar to how humans do~\cite{adams2018naturally}. 

Additionally, we test a modified version of our training loss in Equation~\ref{eq:mae} to take into account the different classes of training labels (strong/weak); we use a loss reweighting scheme~\cite{song2022learning} and define our weighted MAE loss ($L_{wMAE}$) as:
\begin{equation}
 L_{wMAE} = \frac{1}{N} \sum_{j=1}^{N} w(y_{j}) |y_{j} - \hat{y_j}| ,
\end{equation}
where $w(y_{j})$ is a weighting function that depends on the class of the label $y_{j}$. Therefore, our original $L_{MAE}$ loss is equivalent to $L_{wMAE}$ when using a constant weighting function $w(y_{j}) = 1$, which does not differentiate between strong and weak labels.
We show results using different weighting functions in Table~\ref{tab:ablation_w}. As expected, setting higher weights to our weakly labeled images has a negative effect; 
however, assigning higher weights to the strongly labeled images also deteriorates performance.
\nw{We hypothesize that in these latter cases (last two rows of Table~\ref{tab:ablation_w}), the training focuses too much on the strongly labeled images compared to the weakly labeled ones that extend the variability of the training data. Therefore, the models generalize worse, and predictions in our test set (composed of unseen scenes) become less precise. In the limit, excessively high weighting for the strongly labeled data would make it equivalent to using only the \emph{Serrano} dataset. Further research is needed to better understand why the network overfits here to the strongly labeled data.}
\nw{In conclusion,} our simple approach using equal weights, leads to the best results.

\begin{table}
 \caption{Ablation studies by independently changing different design elements of our models: removing data augmentation (w/o aug.) \nw{and} adding background masks (bg. mask).
 \nw{We include results of our weakly supervised gloss predictors, both trained with our BSDF weak labels jointly with the 100\% of the Serrano dataset (S.100\% + BSDF) vs. the 20\% of the Serrano dataset (S.20\% + BSDF)}. We show MAE results on our test dataset.
 }
\centering
\small
\begin{tabular}{c|c|c|c}
 MAE $\downarrow$ & Ours & w/o aug. & bg. mask \\ \hline
 S.100\% + BSDF & \textbf{0.1207} & 0.3113 & 0.1819  \\
 \nw{S.20\% + BSDF} & \textbf{0.1538} & 0.3131 & 0.1976 \\
\hline
\end{tabular}
 \label{tab:ablation}
\end{table}
\begin{table}
\centering
 \caption{Ablation studies by changing the weighting function in our $L_{wMAE}$ training loss. \nw{We include results of our weakly supervised gloss predictors, both trained with our BSDF weak labels jointly with the 100\% of the Serrano dataset (S.100\% + BSDF) vs. the 20\% of the Serrano dataset (S.20\% + BSDF)}. We show MAE results on our test dataset. 
 }
\small
\begin{tabular}{cc|c|c}
\multicolumn{2}{c|}{$w(y_{j})$}  & \multicolumn{2}{c}{MAE $\downarrow$}\\
 strong & weak & S.100\%+BSDF & \nw{S.20\%+BSDF}  \\ \hline \hline
 $1.0$ & $5.0$ & 0.1655 & 0.2685  \\
 $1.0$ & $1.5$ & 0.1634 & 0.1800 \\ \hline
 $1.0$ & $1.0$ &\textbf{0.1207} & \textbf{0.1538} \\ \hline
 $1.5$ & $1.0$ & 0.1361 & 0.1622 \\
 $5.0$ & $1.0$ & 0.1775 & 0.1900 \\
\hline
\end{tabular}
 \label{tab:ablation_w}
\end{table}

\begin{figure}[t!]
   \centering
   \includegraphics[width=0.9\linewidth]{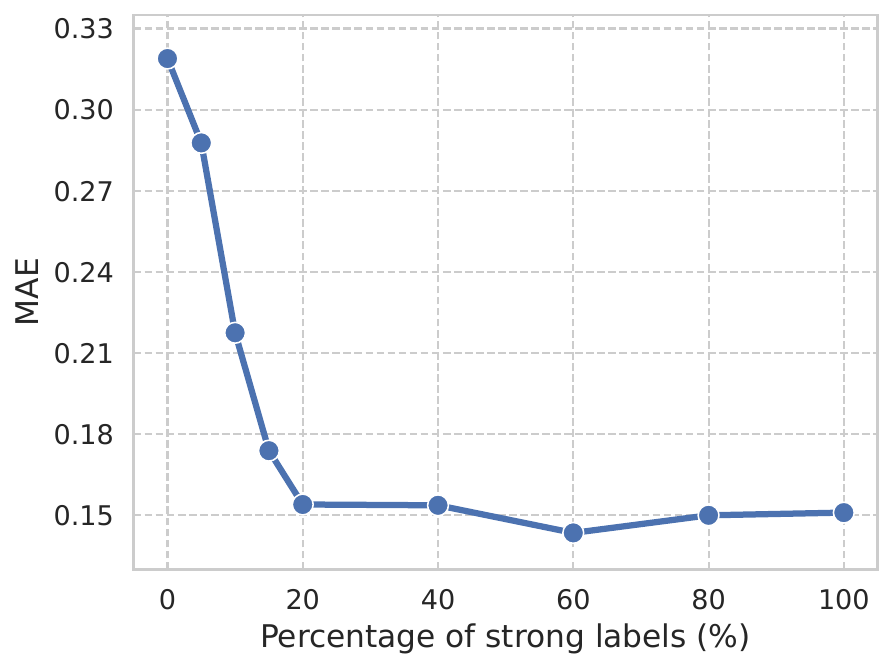}
   \caption{
   \nw{Evolution of the mean absolute error (MAE $\downarrow$) when training our gloss predictor with the same amount of training images but different percentages of strong labels. 0\% indicates that the predictor has been trained with 23,616 randomly selected images from our analytical dataset annotated with BSDF weak labels, while 100\% indicates that the predictor has been trained with all 23,616 images from the Serrano dataset annotated with strong labels. We see that, also when the total number of training images is constant, only a 20\% of strong labels is needed to achieve similar performance to using 100\% of strong labels.
   }
}\label{fig:same_images}
\end{figure}

\nw{Finally, although our weakly-annotated images are almost free to obtain, we validate that the maintained performance is not due to the larger amount of total training images. To do so, we train several models with exactly the same amount of training images (23,616), but varying the percentage of strong labels from the \emph{Serrano} dataset (from 0\% to 100\% = 23,616 strong labels). To complete the training data, we randomly select images from our extended analytical dataset labeled with our BSDF weak labels. 
In Figure~\ref{fig:same_images}, we show that 20\% of strong labels (+80\% of weak labels) are sufficient to achieve a similar performance as using the 100\% of the \emph{Serrano} dataset (100\% of strong labels), hence showing the effectiveness of the weakly supervised strategy independent of the total number of training images.}

\subsection{Towards a Perceptually Meaningful Latent Space}
\label{subsec:latent_space}
\begin{figure}[t!]
    \centering
    \includegraphics[width=\linewidth]{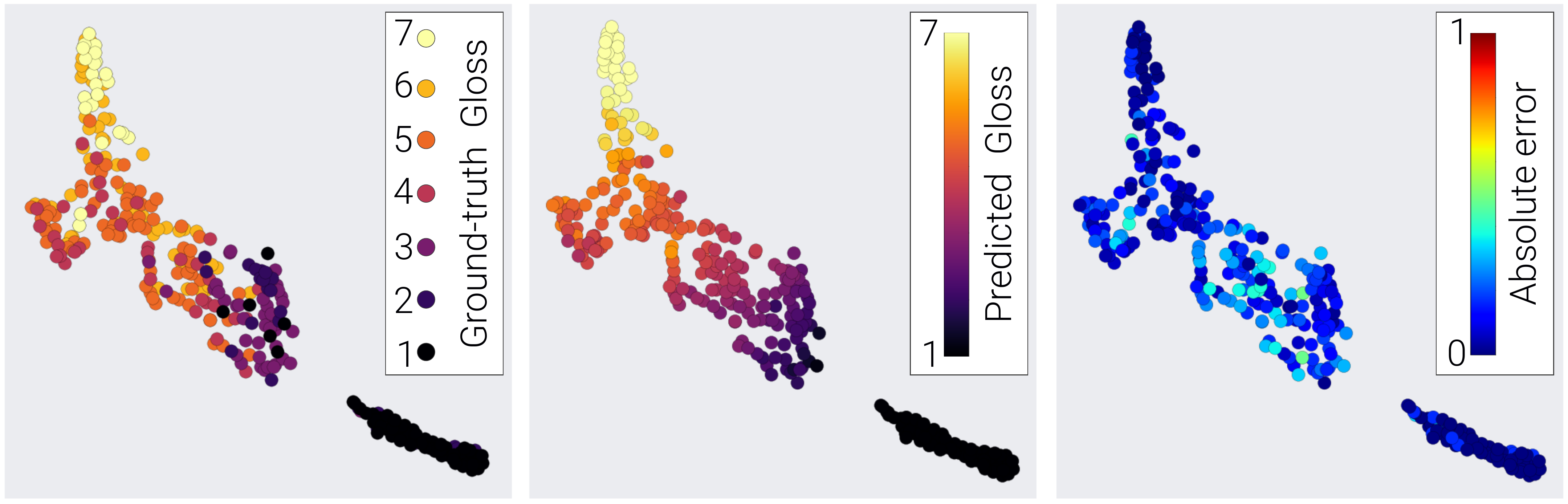}
    \caption{Visualization of the latent space. We show all images from our controlled test dataset as points in 2D space using t-SNE~\cite{tsne} dimensionality reduction from the 20-dimensional feature vectors of the latent space. The color of every point indicates, from left to right: ground-truth gloss judgements, gloss predictions from our model and absolute error (normalized to the range $[0, 1]$). \nw{In this figure, we use our best weakly supervised model, trained on the Serrano dataset and our BSDF weak labels (S.100\%+BSDF).}}
    \label{fig:latent_space}
\end{figure}

Our gloss predictor\nw{s are} designed to encode input images into a low-dimensional latent space $\textbf{z}$ of 20 dimensions. Therefore, if predictions are accurate, this latent space should capture aspects of the underlying structure of gloss perception. As we can see in Figure~\ref{fig:latent_space} (left), the latent space is indeed meaningfully organized with respect to human judgements of gloss for our test dataset. Very matte and very glossy materials (gloss levels 1 and 7) get clearly clustered, likely due to the higher confidence (and therefore consistency) of human labels towards the extreme values of the gloss range. The latent space is also smoothly related to the network's output predicted gloss (Figure~\ref{fig:latent_space}, center), leading to a generally low absolute error (Figure~\ref{fig:latent_space}, right) for all images in the test dataset. This latent space of perceptual gloss could be potentially leveraged for several applications such as material recommendations or database visualization~\cite{lagunas2019similarity}.

\section{Discussion and Future Work}

This paper explores weakly supervised learning in the context of gloss prediction from images. Our work shows that, when modeling perceived material appearance, it may be possible to reduce the high cost of collecting human annotations by leveraging automatically computed weak labels for model supervision. These weak labels, combined with a reduced set of strong human labels, lead to accurate gloss predictions.
Furthermore, our predictions are consistent with human perception of gloss for systematic changes in confounding factors that influence appearance, generalize reasonably well to out-of-distribution images, and exhibit an organized latent space with respect to human perception.

\begin{figure}[t!]
    \centering
    \includegraphics[width=0.75\linewidth]{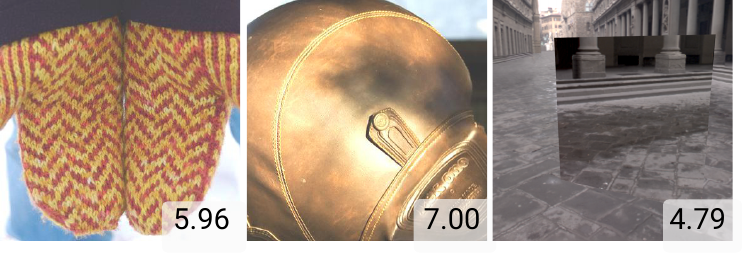}
    \caption{
        Failure cases: real images with multicolored patterned surfaces (left) and very sharp shadows \nw{(center)} are sometimes predicted by our model as highly glossy surfaces; \nw{perfectly-specular flat surfaces are not properly predicted by our model as mirrors (right)}. \nw{To illustrate these general failure cases, we use here our best weakly supervised model (S.100\%+BSDF).} Predictions are in the range $[1, 7]$.
    }
\label{fig:failure}
\end{figure}

Our \nw{work} is not free of limitations. First, although \nw{our weakly supervised gloss predictors} accurately capture the trend of human ratings for systematic variations of appearance, we see how they tend to slightly underestimate the absolute gloss level, especially for \nw{images rendered with} analytical materials. 
Additionally, despite \nw{their} reasonable generalization performance for real images, \nw{our predictions fail} in some challenging cases, such as multicolored patterns, very bright scenes exhibiting sharp shadows \nw{and perfectly-specular flat surfaces}, as shown in Figure~\ref{fig:failure}. \nw{For the first two cases}, we hypothesize this might be due to the network interpreting high-contrast differences as glossy highlights. \nw{For the latter, we believe the network bases its predictions on the materials of the objects reflected on the perfectly specular surface, instead of focusing on the material of the mirror itself, which should be predicted as highly glossy.}
\rev{Providing an object mask as an additional channel could help the gloss predictor to divert its attention from the image background, improving its performance. However, this process can be somewhat cumbersome or imprecise, especially when working with real images.}

Our work inspires promising lines of future research. 
First, we have shown how our simple weakly supervised strategy, using a constant weight that does not take into account the type of label, suffices for effective learning of accurate gloss predictions. \nw{It is not straightforward that weakly supervised learning could be successful in the field of gloss perception, nor what the weak labels should be.} 
\nw{In the future}, we believe that other weakly supervised learning frameworks could be exploited, including, for instance, assigning loss weights automatically based on the noise of each weak label~\cite{ren2018learning, cheng2020weakly}, or transfer learning strategies between weak and strong labels~\cite{robinson2020strength}. 
We have proposed three \nw{one-dimensional} weak labels, conceptually simple and inexpensive to compute.
\nw{Using one-dimensional metrics is inspired by common practice in the literature that operationally defines gloss as a single judgement provided by human observers in response to "how glossy is the surface?"~\cite{serrano2021effect, kim2012dark, marlow12sciencedir}, that captures their overall impression. However, perceived gloss is multi-dimensional~\cite{chadwick15VisRes}, so we believe future work could extend our approach to model multiple dimensions of gloss by defining alternative weak labels and asking observers to rate images in terms of multiple features.}
\rev{On the other hand, although our predictions are consistent with human perception of gloss, we evaluate our predictors on our controlled test dataset, which could be biased towards expert knowledge. Therefore, our work could be extended to evaluate whether our predictions are also consistent with the non-expert human perception of gloss, by collecting a set of test data annotated by naive observers to test whether this yields the same conclusions.
}
\nw{In addition, our datasets (as well as the \emph{Serrano} dataset) contain only opaque materials. We believe the current generalization limitations could be mitigated by augmenting the training data to a wider variety of optical characteristics, such as translucent or iridescent materials, as well as more diverse images (e.g., real photographs or patterned surfaces).}

Finally, our weakly supervised predictors encode images into a low-dimensional latent space that is well organized with respect to perceptual gloss. Exciting future work opens up to explore how to further guide this space in order to disentangle other material appearance factors.

\section*{Acknowledgments}
This work has received funding from the Spanish Agencia Estatal de Investigación (Project PID2022-141539NB-I00 funded by MCIN/AEI/10.13039/501100011033/FEDER, EU) and from the European Union’s Horizon 2020 research and innovation programme under the Marie Skłodowska-Curie grant agreement No. 956585 (PRIME).
This research was also supported by the Deutsche Forschungsgemeinschaft (DFG, German Research Foundation—project number 222641018—SFB/TRR 135 TP C1), by the European Research Council (ERC, Advanced Grant ``STUFF''—project number ERC-2022-AdG-101098225), by the Marsden Fund of the Royal Society of New Zealand (MFP-UOA2109) and by the Cluster Project ``The Adaptive Mind'', funded by the Excellence Program of the Hessian Ministry of Higher Education, Science, Research and Art. 
Julia Guerrero-Viu was supported by the FPU20/02340 predoctoral grant and J. Daniel Subias was supported by the CUS/702/2022 predoctoral grant. 
We also thank Daniel Martin for helping with figures and the members of the Graphics and Imaging Lab for insightful discussions.


\bibliographystyle{Source/eg-alpha-doi}  
\bibliography{bibliography}        
              
\end{document}